\documentclass{PoS}
\usepackage{epsf}
\usepackage{amsmath}
\usepackage{amssymb}
\usepackage{amssymb}
\usepackage{makeidx}
\usepackage{graphicx}
\usepackage{bm}

\newcommand{\betrag}[1]{\left\vert#1\right\vert}

\def\g{\kappa}

\def\x{{\mathbf x}} 

\def\im{{\rm i}}

\def\ie{{\em i.e.\/}}
 
\def\bnabla{\mbox{\boldmath$\nabla$}} 
\def\Box{\kern0.5pt{\lower0.1pt\vbox{\hrule height.5pt width 6.8pt     
\hbox{\vrule width.5pt height6pt \kern6pt \vrule width.3pt}     
\hrule height.3pt width 6.8pt} }\kern1.5pt} 
 
\def\g{{\mbox{\sl g}}}%
\def\Box{\nabla^2}%
\def\i{{\mathrm i}}%
\def\ie{{\em i.e.\/}}%
\def\eg{{\em e.g.\/}}%
\title{Quantum gravity phenomenology via Lorentz violations}

\ShortTitle{Quantum gravity phenomenology}

\author{\speaker{Stefano Liberati}\thanks{liberati@sissa.it}\\
       SISSA/ISAS and INFN, Trieste, Italy\\
        E-mail: \email{liberati@sissa.it}}

\abstract{The search for a quantum theory of gravity has been one of the main aims of theoretical physics for many years by now. However the efforts in this direction have been often hampered by the lack of experimental/observational tests able to select among, or at least constrain, the numerous quantum gravity models proposed so far. This situation has changed in the last decade thanks to the realization that some QG inspired violations of Lorentz symmetry could be constrained using current experiments and observations. This study it is not only allowing us to test at higher and higher energies a fundamental symmetry of spacetime but it is also providing us with hints and perspectives about the fundamental nature of gravity.}

\FullConference{School on Particle Physics, Gravity and Cosmology\\
		21 August - 2 September 2006 \\
		Dubrovnik, Croatia}

\begin{document}

\clearpage%

%

\section{Introduction}

The quest for a quantum theory of gravity has been one of the main enterprise of modern physics for more than half a century. However this research has been systematically frustrated by the lack of experimental/observational test of the models so far proposed. 

Albeit many intriguing and ingenious ideas have been explored, it seems safe to say that without both observing phenomena that depend on Quantum Gravity (QG), and extracting reliable predictions from candidate theories that can be compared with observations, the goal of a theory capable of incorporating quantum mechanics and general relativity will remain unattainable. 

Of course one can claim that a primary tests to be passed by any candidate QG theory is to admit a suitable semiclassical limit, {\em i.e.}~to recover GR at sufficiently low energies. However this is is more a consistency requirement than a true prediction to be tested for a given QG model. In this sense the only example of a prediction of a quantum gravity model that has received some support from observation is the 
spectrum of primordial cosmological perturbations where the quantized longitudinal linearized gravitational mode, albeit slave to the inflaton and not a
dynamically independent degree of freedom, plays an essential role~\cite{Mukhanov:1990me}. However also in this case one might object that somehow any theory of quantum gravity that admits a GR limit should admit as well a regime where gravitons are a valid concept.

Looking for empirical evidence the last decades have witnessed an increasing number of ideas about observable phenomena where QG could play a key role.
A partial list includes: deviations from Newton's law at very short distances \cite{Hoyle:2004cw, Hoyle:2000cv}, Planck-scale fuzziness of spacetime \cite{Amelino-Camelia:1999gg}, possible production in TeV-scale QG scenarios of mini-black holes at colliders \cite{Dimopoulos:2001hw} or in cosmic rays~\cite{Anchordoqui:2001cg}, QG induced violations of discrete symmetries of the Standard Model \cite{Ellis:1995xd,Kostelecky:2005mj} as well as spacetime symmetries \cite{Garay:1994en}. This broad field of research goes under the general name of quantum gravity phenomenology.

We shall focus here on this last item, more specifically on the possibility that the Lorentz symmetry could be violated or deformed (we shall explain later what we exactly mean by this) by Planck-suppressed corrections. Given the necessary conciseness of these proceedings we shall not cover all the related issues, we direct the reader interested in deepening this subject to some of the extensive reviews available in the literature (see {\em e.g.}~\cite{Mattingly:2005re}). Discussion and results presented in this proceeding can be also be found (in an extended form) in the following works \cite{Jacobson:2005bg,Analogues,LSV,Meas2,Meas3,Girelli:2006fw,An-QG}.

\section{Lorentz symmetry at the Planck scale}

Lorentz symmetry has been confirmed to ever
greater precision, and it powerfully constrains theories in a way
that has proved instrumental in discovering new laws of physics.
It seems then natural to assume under these
circumstances that Lorentz invariance is a symmetry of nature up to
arbitrary boosts.  Nevertheless, from a purely logical point of
view, it is clear that such a conjecture is empirically not strongly motivated given that an infinite volume of the Lorentz group is (and will always be) experimentally untested being the Lorentz group non-compact, unlike the rotation group. 
Why should we assume that {\it exact} Lorentz invariance holds at all scales when this hypothesis cannot even in principle be tested? In this sense high energy tests of Lorentz symmetry are ``per se'' valuable being significant improvements of the precision with which an apparently fundamental symmetry of nature is tested.

While the above reasoning is logically sound, it is by itself not very encouraging as it is not providing any argument by which we should expect any departure from Lorentz symmetry due to QG. However, there are also several arguments that lead to suspect that there could be a failure of Lorentz symmetry in proximity of the Planck scale such as the ubiquity of UV divergences in quantum field theories or profound implications related to the very construction of a quantum theory of spacetime (see {\em e.g.}~\cite{Isham,Kuchar,garay}).

Aside from general issues of principle, specific hints of Lorentz
violation have come from  tentative calculations in various approaches to QG hinting that this might be the case. Just to cite a few, we can recall string theory tensor
VEVs~\cite{KS89}, spacetime foam~\cite{GAC-Nat}, semiclassical spin-network calculations in Loop QG~\cite{LoopQG}, non-commutative geometry~\cite{Carroll:2001ws}, some brane-world backgrounds~\cite{Burgess:2002tb}, and condensed matter analogues of ``emergent gravity''~\cite{Analogues}.

Of course, lacking a definitive theory of QG one cannot claim that there is a convincing prediction that some departure from  Lorentz symmetry must be a feature of quantum gravity. However, taken together they do motivate the effort to characterize possible observable consequences of LV and to strengthen observational bounds. Moreover, although very different, these models have in common the fact that Lorentz violations express themselves through modified dispersion relations for elementary particles.
Generically one can cast them in the form:
\begin{equation}%
E^2=p^2+m^2+f(E,p;\mu;M)\;,%
\label{eq:disprel}%
\end{equation}%
where, $c=1$, $E,p$ are the energy and momentum of the particle, $\mu$ is some particle physics mass scale (possibly the mass of the particle $m$) and $M$ denotes the mass scale at which the quantum gravity corrections become appreciable.  Normally, one assumes that $M$ is of order the Planck mass: $M \sim M_{\rm P} \approx 1.22\times 10^{19}\;$GeV. 

We can expand the above dispersion relation in powers of the particle momentum\footnote{We assume, for simplicity, that rotational invariance is preserved and only boost invariance is affected by Planck scale corrections.}
\begin{equation}%
E^2=p^2+m^2+\tilde{\eta}^{(1)} p+\tilde{\eta}^{(2)} p^2+\tilde{\eta}^{(3)} p^3+\tilde{\eta}^{(4)} p^4+\dots\;.%
\label{eq:disprel2}%
\end{equation}%
The dependence on the mass terms $\mu$ and $M$  is now hidden in the dimensionfull coefficients $\tilde{\eta}^{(n)}$. In general the idea that the above dispersion relations are a sort of mesoscopic effect of some Planckian physics leads to the natural expectation that the extra term in Eq.~(\ref{eq:disprel2}) contains suitable powers of the mass terms that make sufficiently Planck suppressed for being compatible with low energy observations.  For the moment we simply follow the observational lead and insert
at least one inverse power of $M$ in each term
\begin{equation} \tilde{\eta}^{(1)}=\eta^{(1)} \frac{\mu^2}{M},\qquad
\tilde{\eta}^{(2)}=\eta^{(2)} \frac{\mu}{M},\qquad \tilde{\eta}^{(3)}=
\frac{\eta^{(3)}}{M},\qquad \tilde{\eta}^{(4)}= \frac{\eta^{(4)}}{M^2}\,.
\label{eq:disprel3}\end{equation}
%
Later on, in characterizing the strength of a constraint we refer to the
$\eta_n$ without the tilde, so we are comparing to what might be
expected from Planck-suppressed LV. We shall also allow the LV parameters
 to dependent on the particle type $\eta^{(n)}=\eta_i^{(n)}$.

It is remarkable that several significant constraints can be put on the intensity of the Lorentz violating term $f(E,p;M)$ using current experiments and observations~\cite{Mattingly:2005re}. However before discussing some of such constraints we  want to provide here an explicit toy model of a system which does show modified dispersion relations of the sort we just conjecture and which for this reason could in fact provide a test field for our ideas. Such a framework is provided by the so called analogue models of gravity~\cite{Analogues}.

\section{Analogue models of gravity}

Reduced to the bones most of the analogue models of gravity can be described as condensed matter systems where the dynamics of atoms or molecule provides an emergent space-time geometry for the propagation of collective excitations of the background. In this sense analogue models of gravity are indeed analogue models of emergent gravity frameworks, {\em i.e.} of scenarios in which the metric and the affine connections are collective variables emerging from the dynamics of more fundamental objects (note that in this sense the metric or the connections would not be the right object to quantize the same way you would not expect to recover atomic interactions from the quantization of hydrodynamical variables).

The simplest example of an analog model is provided by a barotropic --- $\rho=f(p)$, with $\rho$ and $p$ respectively equal to the density and pressure of the fluid --- and inviscid fluid whose flow is irrotational (though possibly time dependent). For such a system the  equation of motion of the velocity potential of an acoustic disturbance $\phi$ (defined as $\mathbf{v} = -{\bnabla} \phi$, where $\mathbf{v}$ is the velocity of the flow)  is identical to the d'Alembertian equation of motion for a minimally coupled massless scalar field propagating in a $(3+1)$--dimensional Lorentzian
geometry
\begin{equation}
\Delta \phi \equiv 
{1\over\sqrt{-g}} 
\partial_\mu 
\left( \sqrt{-g} \; g^{\mu\nu} \; \partial_\nu \phi \right) = 0.
\end{equation}
This signifies that under the above conditions, the propagation of sound is governed by an {\em acoustic metric} --- $g_{\mu\nu}(t,\x)$. This acoustic
metric describes a $(3+1)$--dimensional Lorentzian (pseudo--Riemannian)
geometry. The metric depends algebraically on the density, velocity
of flow, and local speed of sound in the fluid. Specifically
\begin{equation}
g_{\mu\nu}(t,\x) 
\equiv {\rho\over c_{\rm s}} 
\begin{bmatrix}
   -(c_{\rm s}^2-v^2)&\vdots&-{\mathbf v}^T\\
   \cdots\cdots\cdots\cdots&\cdot&\cdots\cdots\\
   -{\mathbf v}&\vdots& \mathbf{I}\\ 
\end{bmatrix},
\label{eq:ac-geom}
\end{equation}
where $c{\rm s}$ is the speed of sound.

Correspondingly the dispersion relation for the quanta of sound (which can be meaningfully defined in some coherent systems like {\em e.g.}~superfluids), the so caled ``phonons'', will take the standard relativistic form 
\begin{equation}
\omega^2=c^2_{\rm s} k^2
\end{equation}
where the role of the speed of light is now played by the speed of sound.

The interesting point for our discussion here is that in a realistic condensed matter system the underlying microscopic structure of the background ({\em e.g.}~the fact that the fluid is made up of molecule) will generically show up in a breakdown of the acoustic Lorentz invariance of the phonon equation of motions which will lead to modified dispersion relations of the form (\ref{eq:disprel2}) where now $M$ will correspond to the energy scale that marks the transition from an hydrodynamic description to a molecular/atomic one ({\em e.g.}~ inter-molecular distance or coherence length). In order to explicitly see this we have however to fix a particular analogue model and a well known example in this sense is that of a Bose-Einstein condensate~\cite{Analogues,Garay:1999sk,Garay:2000jj,Barcelo:2000tg,Visser:2001jd}.


\paragraph{Bose--Einstein condensates:}

Let us start by very briefly reviewing the derivation of the acoustic
metric for a BEC system, and show that the equations for the phonons
of the condensate closely mimic the dynamics of a scalar field in a
curved spacetime.  In the dilute gas approximation, one can describe a
Bose gas through a quantum field ${\widehat \Psi}$ satisfying 
\begin{eqnarray} \im
\hbar \; \frac{\partial }{\partial t} {\widehat \Psi}= \left( -
{\hbar^2 \over 2m} \nabla^2 + V_{\rm ext}(\x) +\g(a)\;{\widehat
\Psi}^{\dagger}{\widehat \Psi} \right){\widehat \Psi}. 
\label{eq:nlscheq}
\end{eqnarray}
Here $\g$ parameterizes the strength of the interactions between the
different bosons in the gas. It can be re-expressed in terms of the
scattering length as 
\begin{eqnarray} \g(a) = {4\pi a \hbar^2\over
m}. 
\end{eqnarray} 
The quantum field can be separated into
a macroscopic (classical) condensate and a fluctuation: ${\widehat
\Psi}=\psi+{\widehat \varphi}$, with $\langle {\widehat \Psi}
\rangle=\psi $. After some suitable approximations (basically neglecting the back-reaction of the excitations on the background, see {\em e.g.}~\cite{Analogues,Barcelo:2003wu}), and adopting the so called  Madelung representation for the wave function of the condensate
\begin{equation} 
\psi(t,\x)=\sqrt{n_c(t,\x)} \; \exp[-{\rm i}\theta(t,\x)/\hbar],  
\end{equation} 
the equation describing the evolution of the background (classical) field $\psi$  (Gross--Pitaevskii equation) can be rewritten as a continuity equation plus an Euler equation:
\begin{eqnarray} 
&& \frac{\partial}{\partial t}n_c+\bnabla\cdot({n_c {\mathbf v}})=0, 
\label{E:continuity}\\ 
&& m\frac{\partial}{\partial t}{\mathbf v}+\bnabla\left(\frac{mv^2}{2}+ 
V_\mathrm{ext}(t,\x)+\g n_c- \frac{\hbar^2}{2m}
\frac{\nabla^{2}\sqrt{n_c}}{\sqrt{n_c}} \right)=0, 
\label{E:Euler1} 
\end{eqnarray} 
where we have defined the irrotational ``velocity field'' by ${\mathbf v}\equiv
{\bnabla\theta}/{m}$.

Let us now note that these equations are completely equivalent to those of an irrotational
and inviscid fluid apart from the existence of the so-called quantum
potential 
\begin{equation}
V_{\rm quantum}=
-\hbar^2\nabla^{2}\sqrt{n_c}/(2m\sqrt{n_c}),
\end{equation}
From what we previously saw it should then be clear that if one neglects the quantum potential then the propagation of acoustic perturbations on the BEC background will be again characterized by an acoustic geometry of the form of (\ref{eq:ac-geom}).

More precisely if one defines $\widehat \theta_1$ as a quantum excitation of the phase of the background $\theta$ (which here plays the role of the velocity potential in the perfect fluid) then one finds that  the equation for the field $\widehat \theta_1$
becomes that of a (massless minimally coupled) quantum scalar field
over a curved background 
\begin{equation}
\Delta\theta_{1}\equiv\frac{1}{\sqrt{-g}}\;
\partial_{\mu}\left(\sqrt{-g}\; g^{\mu\nu}\; \partial_{\nu}\right)
\widehat\theta_{1}=0, 
\end{equation} 
with an effective metric of the
form 
\begin{equation} g_{\mu\nu}(t,\x) \equiv {n_c\over m\;
c_{\rm s}(a,n_c)}
\begin{bmatrix}
   -\{c_{\rm s}(a,n_c)^2-v^2\}&\vdots& - {\rm v}_j \\
   \cdots\cdots\cdots\cdots&\cdot&\cdots\cdots\\
   -{\rm v}_i&\vdots&\delta_{ij}\\
\end{bmatrix}, 
\end{equation} 
where the magnitude
$c_{\rm s}(n_c,a)$ represents the speed of the phonons in the medium: 
\begin{equation} 
c_{\rm s}(a,n_c)^2={\g(a) \; n_c \over m}. 
\end{equation} 

Let us now consider the case in which the above
``hydrodynamical'' approximation for BECs does not hold. In order to
explore a regime where the contribution of the quantum potential
cannot be neglected we can use the so called {\emph{eikonal}}
approximation, a high-momentum approximation where the phase
fluctuation $\widehat \theta_1$ is itself treated as a slowly-varying
amplitude times a rapidly varying phase (see {\em e.g.}~\cite{Analogues,Barcelo:2003wu} for further details). 
Specifically, we shall write 
${\widehat\theta}_1(t,{\mathbf x}) = {\rm Re}\left\{
{\mathcal A}_\theta \; \exp(-i\phi) \right\}$.
As a consequence of  this assumption, gradients  of    the
amplitude, and gradients of  the background fields, are systematically
ignored relative to gradients of $\phi$.  Then adopting the notation  
\begin{equation}  
\omega ={\partial\phi\over\partial t};
\qquad k_i = \nabla_i \phi, \end{equation} 
one can show that the dispersion relation for the excitations of the BEC (quasi-particles) in a flat background ({\em i.e.} when the fluid is at rest, ${\rm v}_0=0$) takes the form: 
\begin{equation} 
\omega^2= c_{\rm s}^2 k^2+c_{\rm s}^2\left({\hbar \over 2 m c_{\rm s}}\right)^2 \;k^4,
\label{eq:disprel-BEC} 
\end{equation} 
which is just a special case of the general form of dispersion relations (\ref{eq:disprel2}) we consider in this work and corresponds to the so called Bogoliubov dispersion relation which was found in 1947~\cite{Bogoliubov:1947} via a diagonalization
procedure for the Hamiltonian describing the system of bosons . Note also that the ``quantum gravity scale'' or more correctly the scale of the violation of the acoustic Lorentz invariance is  the wavelength
$\lambda= 2\pi / ||k||$ with respect to the ``acoustic Compton
wavelength'' $\lambda_c=h/(m c_{\rm s})$.
In particular,  for large wavelengths $\lambda \gg \lambda_c$ one gets a standard
phonon dispersion relation $\omega \approx c ||k||$. For wavelengths
$\lambda \ll \lambda_c$ the quasi-particle energy tends to the kinetic
energy of an individual gas particle and in fact $\omega \approx
\hbar^2 k^2/(2 m)$ (as it should be given that the BEC is in the end a system of many interacting non-relativistic atoms as described by the non-linear Sch\"rodinger equation (\ref{eq:nlscheq}) ).  

The above discussion clearly shows the potentiality of the analogues systems as an inspirational tool for understanding the interplay of micro-physics and macro-physics in quantum gravity phenomenology of Lorentz violations, where the modified dispersion relations of the form (\ref{eq:disprel2}) can be considered intrinsically a mesoscopic physics effect relevant at intermediate energies between our low energy world and the Planck scale. Of course one has to keep in mind that while the relativistic (low energy w.r.t.~$M$) limit of the dispersion relation above is quite generic in many models (the microscopic parameters will in general only determine the value of the speed of sound $c_{\rm s}$), the ``high energy'' deviations ({\em i.e.}~the phenomenology at energies comparable to M) will not be generic but instead strongly dependent on the microphysics of the system. Nonetheless we shall see later on that sometimes useful lessons about viable mechanisms in nature for generating the dispersion relations under investigations can be learn and possibly exported to the more realistic scenarios.

\section{Modified dispersion relations}

Let us now move back to the problem of actually placing constraints on the departures from Lorentz invariance manifested in dispersion relations like (\ref{eq:disprel2}). It was perhaps one of the greatest achievements of the last decade the realization that such constraints can be cast (see e.g.~\cite{Mattingly:2005re} for an extensive review). In fact it was realized that in some special situations even tiny corrections, like the one considered here, can be magnified to observable effects.  A partial list  of these ``windows on quantum gravity'' includes:
\begin{itemize}
\item sidereal variation of Lorentz violation (LV) couplings as the lab moves
  with respect to a preferred frame or directions
\item cumulative effects: long baseline dispersion and vacuum birefringence
  (e.g.~of signals from gamma ray bursts, active galactic
  nuclei, pulsars)
\item anomalous  (normally forbidden) threshold reactions allowed by LV terms (e.g.~photon decay, vacuum
\v{C}erenkov effect) 
\item shifting of existing threshold reactions (e.g.~photon annihilation from
  blazars, GZK reaction)
\item LV induced decays not characterized by a
threshold (e.g.~decay of a particle from one helicity to the other
or photon splitting)
\item maximum velocity (e.g.~synchrotron peak from supernova
remnants)
\item dynamical effects of LV background fields (e.g.
  gravitational coupling and additional wave modes)
\end{itemize}

It has however to be stressed that not all of the above cited tests are in the same way robust against the underlying physical framework that one is choosing in order to justify the use of the modified dispersion relations (\ref{eq:disprel3}). In fact while the above cited cumulative effects use exclusively the form of the modified dispersion relations basically all the other effects are dependent on the underlying dynamics of interacting particles and on the  fact that the standard energy-momentum conservation holds or not. Hence in order to cast most of the constrains on dispersion relations of the form (\ref{eq:disprel2}) one needs to adopt a specific theoretical framework justifying  the use of such deformed dispersion relations.  

\section{Theoretical frameworks for Lorentz violation}

\paragraph{DSR:} 
Quantum gravity effects seems to introduce a new dimensional fundamental scale given by the Planck length (or the corresponding area/volume).  Deformed or Doubly Special  Relativity (DSR) \cite{AC1,ms,AC3} can be
understood as a tentative modification of Special Relativity  (SR) in order
to incorporate some QG scale (generally the Planck length) as a new invariant scale other than that provided by the
speed  of light $c$ but preserving at the same time the relativity principle.   

This very ambitious program has so far found a partial realization only in momentum space where the DSR idea has been implemented by introducing a deformation of the Poincar\'e algebra  in  the boost  sector  \cite{ms,luki}. Specifically  the
Lorentz commutators  among rotations and boost are  left unchanged but
the action of  boosts on momenta is changed in  a non-trivial way (see
e.g.~\cite{ms}) by corrections  which are suppressed by $M$.

Concretely one considers the standard Lorentz algebra of the generators of rotations, $L_i$,
and boosts, $B_i$:%
\begin{equation}%
[L_i,L_j] = \i\,\epsilon_{ijk}\; L_k\;; \qquad%
[L_i,B_j] = \i\,\epsilon_{ijk}\; B_k\;; \qquad%
[B_i,B_j] = -\i\,\epsilon_{ijk}\; L_k%
\label{LB}
\end{equation}%
(Latin indices $i,j,\ldots$ run from 1 to 3) and supplements this with
the following (deformed) commutators between the Lorentz generators and those of
translations in spacetime (the momentum operators $P_0$ and $P_i$):%
\begin{equation}%
[L_i,P_0]=0\;; \qquad%
[L_i, P_j] = \i\,\epsilon_{ijk}\; P_k\;;%
\label{LP}%
\end{equation}%
\begin{equation}%
[B_i,P_0] = \i\; f_1\left({P\over M}\right) P_i\;;%
\label{BP0}%
\end{equation}%
\begin{equation}%
[B_i,P_j] = \i\left[ \delta_{ij} \; f_2\left({P\over M}\right)
P_0 + f_3\left({P\over M}\right)  {P_i \; P_j\over M}
\right]\;.%
\label{BPj}%
\end{equation}%
Finally, assume%
\begin{equation}%
[P_i,P_j] = 0\;.%
\label{PP}%
\end{equation}%
The commutation relations (\ref{BP0})--(\ref{BPj}) are given in terms
of three unspecified, dimensionless structure functions $f_1$, $f_2$,
and $f_3$, and are sufficiently general to include all known DSR
proposals --- the DSR1~\cite{AC1}, DSR2~\cite{ms}, and
DSR3~\cite{AC3}.  Furthermore, in all the DSRs considered to date, the
dimensionless arguments of these functions are specialized to%
\begin{equation}%
f_i\left({P\over M}\right)  \to f_i\left( {P_0\over M},
{\sum_{i=1}^3 P_i^2\over M^2}\right)\;,%
\end{equation}%
so that rotational symmetry is completely unaffected.  Furthermore, in order to recover ordinary special relativity in the $M\to +\infty$ limit, one has to
demand that, in that limit, $f_1$ and $f_2$ tend to $1$, and that
$f_3$ tend to some finite value.%

It was soon recognized~\cite{JV} that such deformed boost algebra
amounts to the assertion that  physical energy and momentum of DSR can
be   always  expressed   as  nonlinear   functions  of   a  fictitious
pseudo-momentum $\pi$,  whose components transform  linearly under the
action  of the  Lorentz group. More  precisely one
can assume  the existence  of an invertible  map $\cal F$  between two
momentum spaces: the {\em  classical space} $\cal P$, with coordinates
$\pi_\mu$ where the Lorentz group  acts linearly and the {\em physical
space} $P$, with coordinates $p_\mu$,  where the Lorentz group acts as
the image of its action on  ${\cal P}$.  
Also, ${\cal F}$ must be such
that ${\cal  F}:[\pi_0,\vec{\pi}]\rightarrow M$ for  all elements
on ${\cal P}$ with $|\vec {\pi}|=\infty$ and/or $\pi_0=\infty$.

The main  open issues in this  momentum formulation of DSR  are the so
called multiplicity  and saturation problems. The first  is related to
the fact  that in principle  there are many possible  deformations (an
infinite number, depending on the choice of an energy invariant scale,
three-momentum scale or both  \cite{luki}). This seems to suggest that
the set  of linear transformations (that  is SR) is the  only one that
have a physical  sense. Moreover  the composition  law for energy  and momenta  of DSR, being derived  by imposing a  standard composition law for  the pseudo
four-momenta $\pi_\mu$, is characterized by a saturation at the Planck
scale apparently  in open contrast with the  everyday life observation
of classical objects with transplanckian energies and momenta.

On the other hand,  since DSR is not a formulation of  QG, but gives a
set of transformations  with the typical QG scale,  it might be plausible to
consider it  as a low energy limit  of QG, that is,  as some effective
theory. (Indeed  such point of view  was taken in several  works on the subject~\cite{dsrqclimit,Magueijo:2002xx,Gir}.) 
Along this line of  though, it was recently proposed~\cite{LSV} that DSR could be interpreted  as an effective theory of measurement for high  energy particle  properties.  According  to this  framework, the
relation between ``true''~ energy and  momentum of a particle (the {\em
  classical} variables $\pi$ of DSR) and observed quantities (the {\em
  physical}  variables  $p$ of  DSR)  acquires,  at sufficiently  high
energies,  Planck suppressed  distortions induced  by  quantum gravity
effects.    These   relations   can   be  identified   with   DSR-type
deformations.  In \cite{Meas2,Meas3} it was further argued that this non linear nature
might arise as a  result of the unavoidable averaging  over QG fluctuations
of  the metric around  flat spacetime  which is  required in  order to
properly define energy  and momentum in first place. 

The main observational constraints on  DSR type dispersion relations concerns limits on the delays between arrival times of sharp features of different energies observed in the intensities of radiation from very distant astrophysical sources. Not very much else is available at the moment given that DSR type dispersion relations do not generically show birefringence and that we cannot resort to anomalous threshold reactions as these phenomena are not allowed in DSR (the reason for this being simply that a kinematically forbidden reaction in the ``classical'' variables $\pi_\mu$
cannot be made viable just via a nonlinear redefinition of momenta). There have been attempts to consider constraints provided by shifts of normally
allowed threshold reactions~\cite{Heyman:2003hs}. However for such reactions the possible constraints are strongly dependent not just on kinematical considerations, but also on reaction rates which require some working framework for their derivation, which presumably would take the form of some Effective Field Theory (EFT).

The problem in this case is the present lack of a DSR formulation in configuration space.
In this sense the main stream approach in the DSR community has been oriented towards the possibility that the configuration space description of the just discussed deformed symmetry in momentum space could imply spacetime non-commutativity~\cite{Amelino-Camelia:2000mn, Kowalski-Glikman:2002jr}. While some promising results linking DSR to some special form of spacetime non-commutativity  have been found in 2+1 dimensions~ \cite{Freidel:2003sp}, we are still lacking a consistent physical  picture in 3+1 dimensions. Moreover attempts to develop a quantum field theory associated with different forms of non-commutativity, a much needed step in order to be able to effectively cast phenomenological constraints, led to
highly non-trivial quantum field theories (possibly with problematic features such as IR/UV-mixing~\cite{Amelino-Camelia:2002au}).

An alternative route it was recently proposed in \cite{Girelli:2006fw} where it was shown that any modified dispersion relation of the form (\ref{eq:disprel2}) can be associated with a Finsler geometry~\footnote{Finsler geometry is a generalization of Riemannian geometry: instead
of defining an inner product structure over the tangent bundle, we
define a norm $F$. This norm will be a real function $F(x,v)$ of a
spacetime point $x$ and of a tangent vector $v\in T_x M$, such that
it satisfies the usual norm properties namely $F(x,v)\neq 0$ if $ v \neq \mathbf{0}$,
and $F(x,\lambda v)= |\lambda | F(x,v)$, with $\lambda \in \mathbb{R}$.
The Finsler metric is then defined as
$g_{\mu\nu}(x,v) = \frac{\partial F^2}{2(\partial v^\mu \partial
v^\nu)},
$ or equivalently
$F(x,v) = \sqrt{g_{\mu\nu}(x,v)v^\mu v^\nu}$.}.
 While this approach can be naturally seen as a geometrical description of the violation of Lorentz symmetry (and in fact it naturally arises in analogue models of gravity~\cite{AMF}), it cannot be excluded at the moment that it might fit instead in a DSR-like framework if a non linear realization of the Lorentz algebra can
be found leaving the  Finsler norm invariant \footnote{It was recently shown in \cite{Mignemi:2007gr} that this is not the case for the standard DSR algebra presented at the start of this section: while the equation of motions are indeed invariant the norm is so only modulo a total derivative. It is presently unclear the physical relevance of this point, given that it is not excluded that other realizations of the Lorentz algebra can leave the norm invariant, neither it is clear the observational relevance the Finsler norm once the invariance of the field equations is assured.}.  This seems an interesting perspective which probably deserve further attention in the next future.
  
In conclusion, missing a field theory implementation of the DSR idea we cannot safely pose most of the constraints listed before. We shall then move to consider the alternative route of an EFT with Lorentz violating terms.

\paragraph{EFT with Lorentz violation:} 
EFT has proven very effective and flexible in the past, it produces local
energy and momentum conservation laws, and seems to require for its
applicability just locality and local spacetime translation invariance above some length scale. It describes the standard model and general relativity (which are presumably not fundamental theories), a myriad of condensed matter systems at appropriate
length and energy scales, and even string theory. Furthermore, it is at the moment the only framework within which we can compute reaction rates and in general fully describe the particle dynamics.

For what regards concrete realizations of this framework we can distinguish two main lines of research: one considers EFT with only renormalizable (\ie~mass dimension 3 and 4) LV operators~\cite{Kc,CG}, the other considers instead EFT with non-renormalizable  (\ie~mass dimension 5 and higher) LV operators~\cite{MP,Jacobson:2005bg}.

Most of the research along the first direction has been carried out within the so called (minimal) standard model extension (SME)~\cite{Kc}. It consists of the standard model
of particle physics plus all Lorentz violating renormalizable
operators (\ie~of mass dimension $\le4$) that can be written
without changing the field content or violating the gauge symmetry.
For illustration, to lowest order in the Lorentz violating coefficients the dispersion relations respectively for electrons and photons in the, rotationally invariant, QED sector of the SME are~\cite{Mattingly:2005re,Kc}

\begin{equation}
E^2 =m^2+p^2+\eta^{(1)}_e p+\eta^{(2)}_e p^2 \qquad E^2 = (1+ \eta^{(2)}_\gamma ){p^2}.
\end{equation}
where from here thereafter we shall identify $m$ with the electron mass $m_e\approx 511$ keV.

The alternative approach is to study non-renormalizable operators. All in all we nowadays consider the SM just an effective field theory and in this sense its renormalizability is seen as a consequence of neglecting some higher order operators which are suppressed in some appropriate mass scale. It is a short deviation from orthodoxy to imagine that such non-renormalizable operators can be generated by quantum gravity effects (and hence be naturally suppressed by the Planck mass) and possibly associated to the violation of some fundamental spacetime symmetry like local Lorentz invariance. 

A rigorous study of Lorentz violating EFT in the higher mass dimension
sector was initiated in~\cite{MP} through a 
classification of all LV dimension five operators that can be added to the
QED Lagrangian and are quadratic in the same fields, rotation
invariant, gauge invariant, not reducible to a combination of lower
and/or higher dimension operators using the field equations, and
contribute $p^3$ terms to the dispersion relation. Just three
operators arise and all of the terms violate CPT symmetry as well as Lorentz
invariance~\footnote{Note that while CPT violation implies LIV the opposite is not true \cite{Greenberg:2002uu}.}. In this case the QED sector dispersion relations come to be 
(in the limit of high energy $E\gg m$)
\begin{eqnarray}
\omega_{\pm}^2&=& k^2 \pm{\xi} \frac{k^3}{M}\label{QEDdisp-ph}\\
E_{\pm}^2&=& p^2 + m^2  +\eta_\pm\frac{p^3}{M}.
\label{QEDdisp-el} 
\end{eqnarray}
were we have introduced the simplified notation $\eta_\gamma^{(3)}=\xi$ and $\eta_e^{(3)}=\eta$. 
The subscripts $\pm$ refer to helicity which can be shown to be a good quantum number in the presence of these LV terms~\cite{JLMS}. 
Moreover, $\eta_\pm$ are the LV parameters of the two
electron helicities, those for positrons can be shown to be $\eta^{\rm positron}_\pm=-\eta^{\rm electron}_\mp$~\cite{Jacobson:2005bg,JLMS}. 

This is the framework in which most of the work on astrophysical constraints has been carried out (with or without taking into account explicitly the possible helicity dependence of the coefficients).

\section{Constraints} 

We shall focus now on the QED sector with dimension 5 LV operators described by the dispersion relations Eq.~(\ref{QEDdisp-ph}) and (\ref{QEDdisp-el}). For these dispersion relations it is obvious that high energies are needed in order to cast constraints. How high, can be estimated considering, for example, threshold processes. In these cases sizable deviations start to appear when the LV term in the dispersion relation is of the order of the relevant mass term. For an electron this implies that one can cast a constraint of $O(1)$ on $\eta$ when $p\sim (mM)^{1/3}\approx 10$ TeV. This energy is at the moment beyond the capacity of terrestrial experiments in particle accelerators however it is well within the realm of high energy astrophysics observations. We give here a very dry account, see \eg~~\cite{Mattingly:2005re,Jacobson:2005bg} for a detailed discussion.

\paragraph{Cumulative effects:} 
Constraints based on cumulative effects are not the strongest available but they do have a special status given that, being a purely of kinematical nature, they are the only ones independent on the choice between the above discussed theoretical frameworks. 

They are cast by placing upper limits on the differences in the time of arrival at Earth of photons produced by a distant astrophysical events. 
In fact one can see from Eq.~(\ref{QEDdisp-ph}) that there is a LIV induced time delay given by 
\begin{equation}
 \Delta t = \xi (k_2-k_1)d/M\;,
\end{equation}
which increases with the distance from the source and with the energy difference. 
Using single sources (generally gamma rays bursts (GRB) or active galactic nuclei (AGN)) the typical strength of the current constraints  strength up to order $10^2$ \cite{TOF}. 

Unfortunately these constraints suffer of large systematic errors, because it is hard to state if the photons, at different energies, are produced simultaneously in the source, even for a GRB. This problem might be avoided using the helicity dependence of  (\ref{QEDdisp-ph}). In fact one can consider the velocity difference of the two polarizations at a \textit{single} energy~\cite{JLMS}. This would lead to a constraint at least twice as large as the one arising from energy differences which is also independent of any intrinsic time lag between different energy photons. However would required the simultaneous detection of both photons' polarizations and the possibility to exclude that some spurious helicity dependent mechanism (like for example the crossing of the light path of some birefringent medium) has affected the relative propagation of the two polarizations states.

Nowadays the most robust constraints on $\xi$ based on this mechanism are obtained via a statistical analysis on a large sample of GRBs with known redshifts. Looking at the arrival times of sharp features in the radiation intensity at different energies it is obtained a constraint of order $\xi\leq O(10^3)$~\cite{GRBroboust}. 

The helicity dependence of (\ref{QEDdisp-ph}) also implies that  the direction of linear polarization is rotated through a frequency-dependent angle, due to different phase velocities for opposite helicities. The difference in rotation angle between two wave vectors $k_1$ and $k_2$ after a propagation distance $d$ is
\begin{equation}
 \Delta\theta = \xi(k_2^2-k_1^2)d/2M\;.
\end{equation}

Albeit small this {\it vacuum birefringence} can easily depolarize linearly polarized radiation composed of a spread of frequencies when this travels over astrophysical distances. Hence detection of polarized light from distant sources can cast a constraint on the strength of the LV coefficient $\xi$. A reliable constraint on the dimension five term, $|\xi|\lesssim 2\times10^{-4}$, was deduced in~\cite{GK} using UV light from distant galaxies. The much stronger constraint $|\xi|\lesssim 2\times10^{-15}$ was
derived~\cite{JLMS,Mitro} from the report~\cite{CB03} of a high
degree of polarization of MeV photons from the gamma-ray burst GRB021206. However, the data has been reanalyzed in two different studies and no statistically significant polarization was found~\cite{RF03}.
Finally, using the same methodology, it was very recently derived the strongest up to date constraint by looking at polarization in the optical/UV afterglow of some gamma-ray bursts (GRB 020813 and GRB 021004). This limit is $|\xi| \lesssim 2\times 10^{-7}$~ \cite{Fan:2007zb}. 

\paragraph{Anomalous threshold reactions:} These are typical phenomena allowed just in case of explicit LV however they are not very sensitive to the details of the dynamics given that  the rates, once above threshold, are tremendously strong Thus we could tolerate huge modifications to the matrix element (the dynamics) and still be able to cast a strong constraint just assuming standard energy momentum conservation (\ie~no DSR). 

The reactions used in this case are mainly those reactions related to the basic vertex of QED, \ie~gamma decay $\gamma\to e^\pm \gamma$ and vacuum \v{C}erenkov (VC) $e^\pm\to e^\pm \gamma$. In both the cases the reaction happens so fast above threshold that the particle stops immediately to propagate~\cite{Jacobson:2005bg}~\footnote{Above threshold the decay rate of photon goes as $\Gamma\sim E^2/M$, while the electron rate of energy loss goes as $dE/dt\sim E^3/M$ \cite{Jacobson:2005bg}.}. Hence if photons or electrons are observed to propagate up to some given high energy one can infer that the latter should be still below the threshold for one of the above reactions to happen.

Taking $\xi \simeq 0~$\footnote{Remember that from birefringence  we already know that $\xi \sim 10^{-7}$.}, the photon decay threshold is
\begin{equation}
k_{th} = 6\sqrt{3}|\eta| m^2M\;.
\end{equation}
In \cite{Jacobson:2005bg} $|\eta| \sim 0.2$ was derived by requiring that that 50~TeV $\gamma$-rays measured from the Crab Nebula had to be below threshold. Now, this constraint is somewhat better, $|\eta| \sim 5\times 10^{-2}$, due to the fact that 80~TeV photons have been seen by HEGRA \cite{Aharonian:2004gb}.

For the vacuum \v{C}erenkov (taking again  $\xi \simeq 0$), the threshold energy is given by
\begin{equation}
 p_{th} = (m^2M/2\eta)^{1/3} \simeq 11\,\mbox{TeV}\,\eta^{-1/3}\;.
\end{equation}
Moreover just above threshold, this process has a time scale of the order of $10^{-9}$ s, so it is extremely efficient. The strongest constraint up to now is again cast using the observation of high energy photons from the Crab nebula: the VHE $\gamma$-ray emission of the Crab Nebula is usually interpreted as due to Inverse Compton (IC) scattering of accelerated electrons/positrons onto background photons. These leptons simply would not be able to produce the observed inverse Compton radiation if they would have been above the VC threshold (the VC rate above threshold is much higher than the IC scattering rate in the Crab). 

In \cite{Jacobson:2005bg} it was used the fact that 50~TeV photons are observed from the Crab Nebula to infer that 50 TeV electrons or positrons in at least one helicity state must be propagating in the Crab nebula.  Thus, the bound $\eta \lesssim 10^{-2}$ was deduced for one of the 4 fermion parameters. With the observation of 80~TeV photons by HEGRA \cite{Aharonian:2004gb} this is strengthened to $\eta \lesssim 3\times10^{-3}$. 

\paragraph{Shifting of existing threshold reactions:} Two characteristic reactions of high energy astrophysics have been studied: photon pair production from TeV gamma rays hitting the infrared (IR) and cosmic microwave background (CMB) photons $\gamma\gamma_0\rightarrow e^+e^-$ and the so called GZK reaction which describes the  production of pions via the collision of ultra high energy protons\footnote{Note that we do not have at the moment a fundamental derivation of the dispersion relation for hadrons (\ie~for composite particles) and a form like Eq.~(\ref{QEDdisp-el}) is just assumed.} ($p\geq 5\cdot 10^{19}$ eV) on CMB, $p+\gamma_{\rm CMB}\rightarrow p+\pi^0$ (see \eg~\cite{Kluzniak,Aloisio:2000cm,GAC-Pir,LIV-JLM}). In the presence of Lorentz violating dispersion relations the threshold for these processes is in general shifted. Moreover, it has been noticed~\cite{Kluzniak, LIV-JLM,LIV-JLM-Th}, that in some cases there is an upper threshold beyond which the process does not occur. It is very difficult to cast robust constraints on reactions of this kind, in fact these are strongly dependent on the shape of the primary spectrum of the incident particles, on the backgrounds (which in the case of IR photons is not well known) and on the reaction rates. Moreover, for GZK reaction, we do have even an uncertainty about the very existence of the reaction hence it is not clear how present observations should be used. For all these reasons we do not discuss here the constraints and refer the reader to the reviews~\cite{Mattingly:2005re,Jacobson:2005bg}.

\paragraph{ LV induced decays not characterized by a
threshold:} Characteristic examples in this case are decays of a particle from one helicity to the other or photon splitting. Although they are not characterized by a proper threshold energy they are strongly dependent on the decay rate which of course is totally negligible at low energies. Helicity decay can happen if the positive and negative helicity LV parameters for electrons are unequal. One can easily see~\cite{Jacobson:2005bg} that the rate is non-negligible for a given $\eta$ approximately at the same energy at which the threshold of the vacuum \v{C}erenkov would be located. In this sense one might speak of an effective threshold. Unfortunately, we lack at the moment sufficiently precise observations for casting constraints using this reaction although the situation might change soon (see discussion in~\cite{Jacobson:2005bg}). For what regards photon splitting we are still missing a complete calculation of the rate which would include the  the helicity dependence of the photon dispersion, and the Lorentz violation in
the electron-positron sector. However preliminary calculations seems to be promising casting a bound on the photon coefficient of order $10^{-3}$~\cite{Gelmini:2005gy}. 

\paragraph{Maximum velocity:}  
Synchrotron radiation emission by electrons/positrons cycling in a magnetic field is strongly affected by LIV. In the Lorentz invariant case, as well as in the presence of Lorentz violation~\cite{Crab,Montemayor:2005ka}, most part of the radiation due to an electron of energy $E$ is emitted around a critical frequency
\begin{equation}
 \omega_c = \frac{3}{2}eB\frac{\gamma^3(E)}{E} \;,
\label{eq:omega_sync}
\end{equation}
where $\gamma(E) =1/\sqrt{ 1-v^2(E)/c^2}$, and $v(E)=\partial E/\partial p$ is the electron group velocity. 

The dispersion relation (\ref{QEDdisp-el}) implies that 
electrons (or positrons) with a negative value of $\eta$ will have a 
maximal group velocity smaller than the low energy speed of light. Consequently
there will be for them a maximal synchrotron frequency $\omega_c^{\rm max}$
 that can be produced, regardless the energy of the radiating lepton.
Thus for at least one electron or positron helicity $\omega_c^{\rm
max}$ must be greater than the maximum observed synchrotron emission
frequency $\omega_{\rm obs}$. This yields a constraint which is strongest for a system that has the smallest $B/\omega_{\rm obs}$ ratio~\cite{Crab}. Presently such a system is the Crab nebula, which emits synchrotron radiation up to 100 MeV and has a magnetic field no larger than 0.6 mG in the emitting region. Thus  one can infer that at least one lepton population must have a LIV coefficient greater than $-7\times10^{-8}$ (note this is not {\em per se}  a constraint, see discussion below). 
 
 \paragraph{Combined constraints:}
Putting all together we get the following picture where all the relevant information is given by the Crab nebula.
\begin{figure}[htb]
\centering
\includegraphics[scale = 0.45]{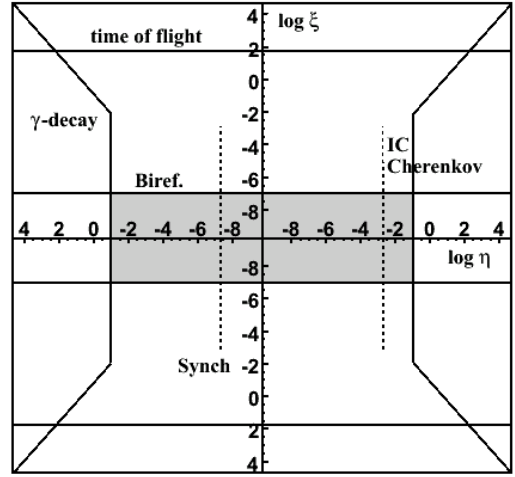}
\caption{Present constraints on the LIV coefficients for QED with dimension 5 Lorentz violation. The grey area is the allowed one. The Dashed lines delimit the allowed range for at least one of the four lepton LIV coefficients if one assumes that a single population has to be simultaneously responsible for the synchrotron and inverse Compton emission of the Crab nebula (from~\cite{MLCK}).}
\label{fig:constraints}
\end{figure}

In summary we see that the $\xi,\eta$ coefficients are nowadays restricted to the region $|\xi|\lesssim10^{-7}$ by birefringence and $|\eta_{\pm}|\lesssim10^{-1}$ by photon decay. It is clear that while the constraint on the photon coefficient is remarkably strong not the same can be said about the LIV coefficients of the leptons. Of course we have a comparably strong estimate on the leptons coefficients, namely the synchrotron one, but this is not a double sided constraint and it only implies that the LIV coefficient of the population responsible for the Crab synchrotron emission cannot be more negative that $-8\times10^{-7}$. The same way the vacuum \v{C}erenkov-IC bound  $\eta<+3\times 10^{-3}$ only tell us that at least one lepton population must satisfy it. These statements, although not void of physical significance, cannot be considered {\em per se} constraints on the parameters $\eta_\pm$, since for each of them it is easy to realize that one of the two parameters $\pm\eta_+$ will always satisfies the bound, as will do one of $\pm\eta_-$. 

Something more can be said making partial use of the information provided by current modeling of the Crab nebula emission. In particular current reconstructions of the Crab emission fit very well the data by just assuming a single lepton population accounting for both the synchrotron and IC emission~\cite{Jacobson:2005bg}. In this case it is possible to infer that at least one of the four pairs $(\pm\eta_\pm,\xi)$ must lie in the narrow region bounded horizontally by the dashed lines of the synch and IC bounds and vertically by the birefringence constraint.   However, we cannot {\em a priori} exclude that only one out of four populations is responsible for both the synchrotron and inverse Compton emission, so the electron sector is not yet strongly constrained (see again \cite{Jacobson:2005bg} for further details).

Nonetheless, it is clear that these simple arguments do not fully exploit the large amount of information we obtain from the Crab Nebula. A detailed comparison of the observations with the reconstructed spectrum in the LIV case, where all new reactions and modifications of classical processes are considered, could still provide us with strong constraints on both positive and negative $\eta$ for the four lepton populations. This study is currently in progress and there is reasonable expectation that it might settle this issue~\cite{MLCK}.

 \section{The naturalness problem: an analogue models lesson}

Looking back at our ansatz Eq.~(\ref{eq:disprel3}) compatibility with observations induced us to assume that the lowest order coefficients $\eta^{(1)}$ and $\eta^{(2)}$ contain some appropriate power of the small ratio $\mu/M$. Note however that we did not assumed any extra Planck suppression for the higher order dimensionless coefficients ($\eta^{(n)}$ with $n\geq3$) which are then naturally of order one. Such an ansatz assures that  for any $p\gg\mu$ the lowest higher order term allowed is always dominant on both the lowest oder ones as well as on all the other higher order ones. If CPT invariance is not a priori enforced, the dominant term would be the one cubic in the momentum in Eq.~(\ref{eq:disprel2}). 

A naturalness problem arises because such a line of reasoning does not seem to be well justified within an EFT framework. In fact we implicitly assumed that there are no extra Planck suppressions hidden in the dimensionless coefficients $\eta^{(n)}$ with $n\geq 3$. However we cannot justify why \emph{only} the dimensionless coefficients of the $n\leq 2$ terms should be suppressed by powers of the small ratio $\mu/M$.  
Furthermore it is easy to show~\cite{Collins} that, without some protecting symmetry, it is generic that radiative corrections due to particle interactions in an  EFT with only Lorentz violations of order $n\geq 3$ in (\ref{eq:disprel2}) for the free particles, will generate $n=1$ and $n=2$ Lorentz violating terms in the dispersion relation which will then be dominant.  

 It has indeed been suggested in~\cite{Pospelov-Nibbelink} that supersymmetry (SUSY) could play a protective role for the lowest-order operators: the dual requirements of supersymmetry and gauge invariance permit one to add to the SUSY standard model only those operators corresponding to $n\geq3$ terms in the dispersion relation. However SUSY is broken in the real world and when LV SUSY QED with softly broken SUSY was considered~\cite{Bolokhov:2005cj}, it was found that, upon SUSY breaking, the dimension five SUSY operators generate dimension three operators large enough that the dimension five operators must be suppressed by a mass scale much greater than $M$. In this sense, the naturalness problem is
{\it not} completely solved in this setting. Alternatively one can try to identify LV operators that are protected against transmutation in lower order ones by a variety of other mechanisms like the irreducibility of the LV tensor structures,T-invariance, and the lepton number conservation. This has been explicitly done for dimension 5 Lorentz Violating Interactions in the Standard Model showing that some LIV operators are indeed protected in this way~\cite{Bolokhov:2007yc}~\footnote{Renormalization group equations for QED with dimension five LIV terms have been also given in \cite{Bolokhov:2007yc}. Solving such equations can be used to show that  $\eta_{\pm}$ and $\xi$ are of order 1 at the TeV scale~\cite{MLCK}. }.  

Finally also analogue models of gravity can be used as an inspiring tool in tackling the naturalness problem. We already saw the wave equation for phonons propagating on a Bose--Einstein condensate BEC can be described  as that of a relativistic scalar field on a curved spacetime (the background is flat in the special case that the BEC is at rest and has constant density). Moreover the dispersion relations for such phonons (Eq.~(\ref{eq:disprel-BEC})) does indeed show high energy corrections. However, in order to address the naturalness problem one needs to consider a system where there are at least two kind of excitations/quasi-particles which moreover have to share the same Lorentzian geometry at low energies. In this sense a system of two coupled BEC is ideal because does satisfy this criteria when its microscopic quantities (background densities and couplings) are suitably tuned~\cite{An-QG}.

The basis of the model is an ultra-cold dilute atomic gas of $N$ bosons in two coupled single-particle states $|A\rangle$ and $|B\rangle$. There are three atom-atom coupling constants, $U_{AA}$, $U_{BB}$, and $U_{AB}$, and an additional coupling $\lambda$ that drives transitions between the two single-particle states. Ignoring again back reaction effects of the quantum fluctuations one then obtains a pair of coupled Gross--Pitaevskii equations (GPE)
\begin{eqnarray}  \label{2GPE} 
 i \, \hbar \, \partial_{t} \psi_{i} &=& \bigg[
   -\frac{\hbar^2}{2\,m_{i}} \nabla^2 + V_{i}-\mu_{i} 
   + U_{ii}
   \, \betrag{\psi_{i}}^2 + U_{ij} \betrag{\psi_{j}}^2
   \bigg] \psi_{i} 
    + \lambda \, \psi_{j} \, , 
\end{eqnarray}
where  $(i,j)\rightarrow (A,B)$ and again $\psi_i$ identify the classical wave function $ \langle \hat\Psi\rangle$ of each condensate specie. 

Now consider small perturbations (sound waves) in the condensate cloud.  The excitation spectrum is obtained by linearizing around some background, and after a straightforward analysis and a suitable tuning of the microscopic quantities it is possible to show that the dispersion relations at low energies are~\cite{An-QG}  
\begin{eqnarray}
\omega^2_I=c^2_{\rm s} k^2\\
\omega^2_{II}=c^2_{\rm s} k^2+m_{II}^2 c^4_{\rm s}
\end{eqnarray}
where $I$ and $II$ identified the two phon-like modes of the coupled system and where both the low energy speed of sound $c_{\rm s}$ and the effective mass $m$ are a function of the microscopic parameters. In particular $m_{II}\to 0$ if the laser coupling $\lambda\to 0$. 

One can then investigate if at high energies the interaction between these modes allows the primary quartic order (in particle momentum) LV to ``percolate'' to quadratic order (hence inducing non-zero $\eta^{(2)}$ terms) and make the latter the dominant contribution. In~\cite{An-QG} it was shown that at high energies one indeed finds that the high energy dispersion relations become~\cite{An-QG}
\begin{eqnarray}
\omega^2_I=c^2_{\rm s} ( k^2+ \eta^{(4)}_I \, k^4/M_{\rm eff}^2),\\
\omega^2_{II}=m_{II}^2 c^4_{\rm s}+c^2_{\rm s} ( \eta^{(2)}_{II} \, k^2+ \eta^{(4)}_{II} \, k^4/M_{\rm eff}^2 ),
\end{eqnarray}
where $M_{\rm eff}=\sqrt{m_A m_B}$ plays the role of the LIV scale. 

So that a percolation of the standard (for BEC) quartic LIV terms into quadratic ones seems to have taken place (at least for the massive quasi-particle). However looking in detail at the form of the $\eta^{(2)}_{II}$ and $\eta^{(4)}_{I/II}$ coefficients one finds that while the former one is actually a small ratio $\eta^{(2)}_{II}=(m_{II}/M_{\rm eff})^2$~\footnote{One can always tune the microscopic physics of the 2-BEC system so that $m_{II}\ll M_{\rm eff}$} the latter are indeed coefficients of order one. So for any $p\gg m_{II}$ the higher order LIV terms will dominate over the quadratic one.

We can here propose a nice interpretation in terms of  ``emergent symmetry'': Non-zero $\lambda$ \emph{simultaneously} produces a non-zero mass for one of the phonons, \emph{and} a corresponding non-zero LIV at order $k^2$.  Let us now drive $\lambda\to 0$ but in such a ay that at low energies the Lorentzian geometry si still recovered (this can always be done~\cite{An-QG}). In this case one gets an EFT which at low energies describes two non-interacting phonons propagating on a common background (both $m_{II}$ and $\eta^{(2)}$ go to 0 for $\lambda\to 0$). Now this system possesses a $SO(2)$ symmetry and hence non-zero laser coupling $\lambda$ softly breaks this $SO(2)$, the mass degeneracy, and  low-energy Lorentz invariance. Such soft Lorentz violation is then characterized (as usual in EFT) by the ratio of the scale of the symmetry breaking $m_{II}$, and that of the scale originating the LIV in first place $M_{\rm LIV}$. We stress that the $SO(2)$ symmetry is an ``emergent/accidental symmetry'' as it is not preserved beyond the hydrodynamic limit: the $\eta^{(4)}$ coefficients are in general different if $m_A\neq m_B$, so $SO(2)$ is generically broken at high energies.  Nevertheless this is enough for the protection of the {\em lowest}-order LIV operators. 

The interesting lesson to be learn concerns the fact that in this case one can see that the symmetry that protects the lowest order operators (those associated with the $\eta^{(2)}$ term) is not an high energy one eventually broken at low energies, on the contrary it is an emergent (low energy) symmetry which being softly broken by interactions allows only for suppressed low order terms. It would be intriguing to explore the possibility that something along these lines could happen in the real world.

\section{Conclusions}

What we presented here is just a succinct account of an ongoing effort to constraints QG induced departures from Lorentz symmetry at high (but still sub-Planckian) energies. It is remarkable that even without experiments specifically dedicated to the detection of LV effects in astrophysics so much could be done in constraining the EFT framework of~\cite{MP}. In the future, experiments like AUGER, GLAST or OWL could help in improving our knowledge of the microscopic spacetime nature. There is also some hope that high energy neutrino astrophysics might play a pivotal role in the next decade advancing this branch of research. We direct again to~\cite{Jacobson:2005bg} for further discussion. 
Moreover we wanted to stress here the interesting role which the interplay of different fields of physics --- high energy physics, astrophysics, condensed matter physics --- can have in this branch of research. All in all while a theory of quantum/emergent gravity is definitely in the realm of high energy physics, it is also true that our best option to test it is by describing its semiclassical limit via EFT scenarios, {\em i.e.}~the ``heart and soul'' of condensed matter physics, and constrain their prediction using the highest energies observations so far available, i.e. the high energy astrophysics one. Who is writing strongly feels that the new challenges ahead us will not be overcome without further exploiting this rich mixing of ideas. In any case we hope that this brief account of the present achievements might have convinced the reader that the age of purely speculative QG models has ended and that the dawn of quantum gravity phenomenology has finally arrived.

\section*{Acknowledgments}
I wish to thank T.A.~Jacobson and L.~Maccione for useful remarks on the manuscript. Figure \ref{fig:constraints} was produced by L.~Maccione for the manuscript \cite{MLCK} in preparation.


\begin{thebibliography}{99}
\bibitem{Mukhanov:1990me}
V.~F.~Mukhanov, H.~A.~Feldman and R.~H.~Brandenberger, 
Phys.\ Rept.\  {\bf 215}, 203 (1992).


\bibitem{Hoyle:2004cw}
  C.~D.~Hoyle, D.~J.~Kapner, B.~R.~Heckel, E.~G.~Adelberger, J.~H.~Gundlach, U.~Schmidt and H.~E.~Swanson,
  Phys.\ Rev.\ D {\bf 70} (2004) 042004
\bibitem{Hoyle:2000cv}
  C.~D.~Hoyle, U.~Schmidt, B.~R.~Heckel, E.~G.~Adelberger, J.~H.~Gundlach, D.~J.~Kapner and H.~E.~Swanson,
  Phys.\ Rev.\ Lett.\  {\bf 86} (2001) 1418

\bibitem{Amelino-Camelia:1999gg}
  G.~Amelino-Camelia,
  Phys.\ Rev.\ D {\bf 62}, 024015 (2000)
  
\bibitem{Dimopoulos:2001hw}
  S.~Dimopoulos and G.~Landsberg,
  Phys.\ Rev.\ Lett.\  {\bf 87} (2001) 161602

\bibitem{Anchordoqui:2001cg}
  L.~A.~Anchordoqui, J.~L.~Feng, H.~Goldberg and A.~D.~Shapere,
  Phys.\ Rev.\ D {\bf 65}, 124027 (2002)
  
  
\bibitem{Ellis:1995xd}
  J.~R.~Ellis, J.~L.~Lopez, N.~E.~Mavromatos and D.~V.~Nanopoulos,
  Phys.\ Rev.\ D {\bf 53}, 3846 (1996)
  
\bibitem{Kostelecky:2005mj}
  V.~A.~Kostelecky,
  {\em ``CPT and Lorentz symmetry''}. Proceedings of the 3rd Meeting on CPT and Lorentz Symmetry (CPT 04), Bloomington, Indiana, 4-7 Aug 2004. Published in Hackensack, USA: World Scientific (2005).

\bibitem{Garay:1994en}
  L.~J.~Garay,
  Int.\ J.\ Mod.\ Phys.\ A {\bf 10}, 145 (1995)
  
\bibitem{Mattingly:2005re}
D.~Mattingly,
  Living Rev.\ Rel.\  {\bf 8}, 5 (2005)

\bibitem{Jacobson:2005bg}
  T.~Jacobson, S.~Liberati and D.~Mattingly,
  Annals Phys.\  {\bf 321}, 150 (2006)

\bibitem{Analogues}
C.~Barcelo, S.~Liberati and M.~Visser,
 {\em ``Analogue gravity,''}
  Living Rev.\ Rel.\  {\bf 8}, 12 (2005).

\bibitem{LSV}
  S.~Liberati, S.~Sonego and M.~Visser,
  Phys.\ Rev.\ D {\bf 71}, 045001 (2005). 
%

\bibitem{Meas2}
  R.~Aloisio, A.~Galante, A.~Grillo, S.~Liberati, E.~Luzio and F.~Mendez,
  Phys.\ Rev.\ D {\bf 73}, 045020 (2006).

\bibitem{Meas3}
  R.~Aloisio, A.~Galante, A.~F.~Grillo, S.~Liberati, E.~Luzio and F.~Mendez,
  Phys.\ Rev.\  D {\bf 74}, 085017 (2006)
\bibitem{Girelli:2006fw}
  F.~Girelli, S.~Liberati and L.~Sindoni,
  Phys.\ Rev.\  D {\bf 75}, 064015 (2007)
  \bibitem{An-QG}
  S.~Liberati, M.~Visser and S.~Weinfurtner,
  Phys.\ Rev.\ Lett.\  {\bf 96}, 151301 (2006).
%
  Class.\ Quant.\ Grav.\  {\bf 23}, 3129 (2006).

\bibitem{Isham}
C.~J.~Isham, {\em``Structural issues in quantum gravity''},
[arXiv:gr-qc/9510063];
\\
J.~Butterfield and C.~J.~Isham, 
{\em ``On the emergence of time in quantum gravity''}, 
[arXiv:gr-qc/9901024].

\bibitem{Kuchar}
K.~V.~Kuchar, {\em ``Time And Interpretations Of Quantum Gravity''},
In ``Winnipeg 1991, Proceedings, General relativity and relativistic
astrophysics'', 211-314.

\bibitem{garay}
L.~J.~Garay, 
Int.\ J.\ Mod.\ Phys.\ A {\bf 10}, 145 (1995).

\bibitem{KS89}
V.~A.~Kostelecky and S.~Samuel, 
Phys.\ Rev.\ D {\bf 39}, 683 (1989).

\bibitem{GAC-Nat}
  G.~Amelino-Camelia, J.~R.~Ellis, N.~E.~Mavromatos and D.~V.~Nanopoulos,
  Int.\ J.\ Mod.\ Phys.\ A {\bf 12}, 607 (1997);\\
G.~Amelino-Camelia, J.~R.~Ellis, N.~E.~Mavromatos, D.~V.~Nanopoulos
and S.~Sarkar, 
Nature {\bf 393}, 763 (1998). 

\bibitem{LoopQG}
R.~Gambini and J.~Pullin, 
Phys.\ Rev.\ D {\bf 59}, 124021 (1999);
J.~Alfaro, H.~A.~Morales-Tecotl and L.~F.~Urrutia, 
Phys.\ Rev.\ D {\bf 65}, 103509 (2002). 

\bibitem{Carroll:2001ws}
S.~M.~Carroll, J.~A.~Harvey, V.~A.~Kostelecky, C.~D.~Lane and T.~Okamoto, 
Phys.\ Rev.\ Lett.\ {\bf 87}, 141601 (2001).

\bibitem{Burgess:2002tb}
C.~P.~Burgess, J.~Cline, E.~Filotas, J.~Matias and G.~D.~Moore,
JHEP {\bf 0203}, 043 (2002).

\bibitem{Garay:1999sk}
  L.~J.~Garay, J.~R.~Anglin, J.~I.~Cirac and P.~Zoller,
  Phys.\ Rev.\ Lett.\  {\bf 85}, 4643 (2000)
  
\bibitem{Garay:2000jj}
  L.~J.~Garay, J.~R.~Anglin, J.~I.~Cirac and P.~Zoller,
  Phys.\ Rev.\  A {\bf 63}, 023611 (2001)
  
\bibitem{Barcelo:2000tg}
  C.~Barcelo, S.~Liberati and M.~Visser,
  Class.\ Quant.\ Grav.\  {\bf 18}, 1137 (2001)
  
\bibitem{Visser:2001jd}
  M.~Visser, C.~Barcelo and S.~Liberati,
  ``Acoustics in Bose-Einstein condensates as an example of broken Lorentz
  symmetry,''
  [arXiv:hep-th/0109033].

\bibitem{Barcelo:2003wu}
  C.~Barcelo, S.~Liberati and M.~Visser,
  Phys.\ Rev.\  A {\bf 68}, 053613 (2003)
  [arXiv:cond-mat/0307491].
  
\bibitem{Bogoliubov:1947}
N.~Bogoliubov, 
J. Phys. (USSR) {\bf 11}, 23 (1947).

\bibitem{AC1}
G.~Amelino-Camelia, %
Int.\ J.\ Mod.\ Phys.\ D {\bf 11}, 35--59 (2002). %
\bibitem{ms}%
J.~Magueijo and L.~Smolin, %
Phys.\ Rev.\ Lett.\ {\bf 88}, 190403 (2002). %
[arXiv:hep-th/0112090];\\%
scale,'' %
Phys.\ Rev.\ D {\bf 67}, 044017 (2003) %
[arXiv:gr-qc/0207085].%
\bibitem{AC3}%
G.~Amelino-Camelia, %
Int.\ J.\ Mod.\ Phys.\ D {\bf 12}, 1211--1226 (2003) %
[arXiv:astro-ph/0209232].%

\bibitem{luki}
J. Lukierski and A. Nowicki, Int. J. Mod. Phys. {\bf
A18}, 7 (2003).


\bibitem{JV}%
S.~Judes and M.~Visser, %
Phys.\ Rev.\ D {\bf 68}, 045001 (2003). %

\bibitem{dsrqclimit}L. Freidel, J.  Kowalski-Glikman and Lee Smolin,
Phys. Rev. {\bf D69}, 044001 (2004); J.  Kowalski-Glikman, 
{\em  3rd  International  Sakharov  Conference  on  Physics},  Moscow,
Russia, 24-29 Jun 2002 [hep-th/0209264].

\bibitem{Magueijo:2002xx}
  J.~Magueijo and L.~Smolin,
   Class.\ Quant.\ Grav.\  {\bf 21}, 1725 (2004).


\bibitem{Gir}
  F.~Girelli, E.~R.~Livine and D.~Oriti,
  Nucl.\ Phys.\ B {\bf 708}, 411 (2005);
%
  F.~Girelli and E.~R.~Livine, arXiv:gr-qc/0412004]; 
  Braz.\ J.\ Phys.\  {\bf 35}, 432 (2005).
  
\bibitem{Heyman:2003hs}
  D.~Heyman, F.~Hinteleitner and S.~Major,
  Phys.\ Rev.\ D {\bf 69}105016  (2004). 
  
\bibitem{Amelino-Camelia:2000mn}
  G.~Amelino-Camelia,
  Int.\ J.\ Mod.\ Phys.\ D {\bf 11} (2002) 35
  [arXiv:gr-qc/0012051].
  
\bibitem{Kowalski-Glikman:2002jr}
  J.~Kowalski-Glikman and S.~Nowak,
  ``Non-commutative space-time of doubly special relativity theories,''
  Int.\ J.\ Mod.\ Phys.\ D {\bf 12} (2003) 299
  [arXiv:hep-th/0204245].

\bibitem{Freidel:2003sp}
  L.~Freidel, J.~Kowalski-Glikman and L.~Smolin,
  Phys.\ Rev.\ D {\bf 69}, 044001 (2004)
  [arXiv:hep-th/0307085].



\bibitem{Amelino-Camelia:2002au}
  G.~Amelino-Camelia, G.~Mandanici and K.~Yoshida,
  ``On the IR / UV mixing and experimental limits on the parameters of
  canonical noncommutative spacetimes,''
  JHEP {\bf 0401} (2004) 037
  [arXiv:hep-th/0209254].

  
  \bibitem{AMF}
 C.~Barcelo, S.~Liberati and M.~Visser,
  Class.\ Quant.\ Grav.\  {\bf 19}, 2961 (2002)
  [arXiv:gr-qc/0111059];\\
   S.~Weinfurtner, S.~Liberati and M.~Visser,
  [arXiv:gr-qc/0605121].
   
\bibitem{Mignemi:2007gr}
  S.~Mignemi,
  arXiv:0704.1728 [gr-qc].


  \bibitem{Kc}
D.~Colladay and V.~A.~Kostelecky,
Phys.\ Rev.\ D {\bf 55}, 6760 (1997);
\\
Phys.\ Rev.\ D {\bf 58}, 116002 (1998).

\bibitem{CG}
S.~R.~Coleman and S.~L.~Glashow,
Phys.\ Rev.\ D {\bf 59}, 116008 (1999).
 
\bibitem{MP}
R.~C.~Myers and M.~Pospelov, 
Phys.\ Rev.\ Lett.\  {\bf 90}, 211601 (2003).

\bibitem{Greenberg:2002uu}
  O.~W.~Greenberg,
  Phys.\ Rev.\ Lett.\  {\bf 89}, 231602 (2002).
  
\bibitem{JLMS}
T.~A.~Jacobson, S.~Liberati, D.~Mattingly and F.~W.~Stecker, 
Phys. Rev. Lett. {\bf 93}, 021101 (2004)

\bibitem{TOF}
S.~D.~Biller {\it et al.}, ``Limits to quantum gravity effects from
observations of TeV flares in  active galaxies,'' Phys.\ Rev.\
Lett.\  {\bf 83}, 2108 (1999) [arXiv:gr-qc/9810044];
%
B.~E.~Schaefer, 
Phys.\ Rev.\ Lett.\ {\bf 82}, 4964 (1999).
%
P.~Kaaret, ``Pulsar radiation and quantum gravity,'' Astronomy and
Astrophysics, {\bf 345}, L32-L34 (1999) [arXiv:astro-ph/9903464];
%
S.~E.~Boggs, C.~B.~Wunderer, K.~Hurley and W.~Coburn,
Astrophys.\ J.\  {\bf 611}, L77 (2004).

\bibitem{GRBroboust}
  J.~R.~Ellis, N.~E.~Mavromatos, D.~V.~Nanopoulos, A.~S.~Sakharov and E.~K.~G.~Sarkisyan,
  Astropart.\ Phys.\  {\bf 25}, 402 (2006).

\bibitem{GK}
R.~J.~Gleiser and C.~N.~Kozameh, 
Phys.\ Rev.\ D {\bf 64}, 083007 (2001).

\bibitem{Mitro}
I.~G.~Mitrofanov, 
Nature {\bf 426}, 139 (2003).

\bibitem{CB03}
W.~Coburn and S.~E.~Boggs, 
Nature {\bf 423}, 415 (2003).

\bibitem{RF03}
R.~E.~Rutledge and D.~B.~Fox, ``Re-Analysis of Polarization in the
Gamma-ray flux of GRB 021206,'' [arXiv:astro-ph/0310385];
S.~E.~Boggs and W.~Coburn, ``Statistical Uncertainty in the
Re-Analysis of Polarization in GRB021206,''
[arXiv:astro-ph/0310515].
C. Wigger et al., ``Gamma-Ray Burst Polarization: Limits from RHESSI
Measurements,'' [arXiv:astro-ph/0405525].


\bibitem{Fan:2007zb}
  Y.~Z.~Fan, D.~M.~Wei and D.~Xu,
  ``Gamma-ray Burst UV/optical afterglow polarimetry as a probe of Quantum Gravity,''
  [arXiv:astro-ph/0702006].

\bibitem{Aharonian:2004gb}
  F.~Aharonian {\it et al.}  [The HEGRA Collaboration],
  Astrophys.\ J.\  {\bf 614}, 897 (2004)

\bibitem{Kluzniak}
W.~Kluzniak, 
Astropart.\ Phys.\  {\bf 11}, 117 (1999).

\bibitem{Aloisio:2000cm}
  R.~Aloisio, P.~Blasi, P.~L.~Ghia and A.~F.~Grillo,
  Phys.\ Rev.\ D {\bf 62}, 053010 (2000).
  
\bibitem{GAC-Pir}
G.~Amelino-Camelia and T.~Piran, 
Phys.\ Rev.\ D {\bf 64}, 036005 (2001).

\bibitem{LIV-JLM}
  T.~Jacobson, S.~Liberati and D.~Mattingly,
  Phys.\ Rev.\ D {\bf 66}, 081302 (2002);
  Phys.\ Rev.\ D {\bf 67}, 124011 (2003).
    
\bibitem{LIV-JLM-Th}
  D.~Mattingly, T.~Jacobson and S.~Liberati,
  Phys.\ Rev.\ D {\bf 67}, 124012 (2003).

\bibitem{Gelmini:2005gy}
  G.~Gelmini, S.~Nussinov and C.~E.~Yaguna,
  JCAP {\bf 0506}, 012 (2005).


\bibitem{Crab}
  T.~Jacobson, S.~Liberati and D.~Mattingly,
  Nature {\bf 424}, 1019 (2003).


\bibitem{Montemayor:2005ka}
  R.~Montemayor and L.~F.~Urrutia,
  Phys.\ Rev.\  D {\bf 72}, 045018 (2005)
  [arXiv:hep-ph/0505135].
  
  \bibitem{MLCK}
L.~Maccione, S.~Liberati, A.~Celotti, J.~Kirk,\\
``The Crab Nebula constraints on Planck-scale Lorentz Violation in QED'',\\
In preparation

\bibitem{Collins}
  J.~Collins, A.~Perez, D.~Sudarsky, L.~Urrutia and H.~Vucetich,
  Phys.\ Rev.\ Lett.\  {\bf 93}, 191301 (2004).

\bibitem{Pospelov-Nibbelink}  
S.~G.~Nibbelink and M.~Pospelov, 
Phys.\ Rev.\ Lett.\ {\bf 94}, 081601 (2005).
\bibitem{Bolokhov:2005cj}
  P.~A.~Bolokhov, S.~G.~Nibbelink and M.~Pospelov,
  Phys.\ Rev.\ D {\bf 72}, 015013 (2005).


\bibitem{Bolokhov:2007yc}
  P.~A.~Bolokhov and M.~Pospelov,
  ``Classification of dimension 5 Lorentz violating interactions in the
  standard model,''
  [arXiv:hep-ph/0703291].


\end{thebibliography}
\end{document}